\documentclass{ws-ijgmmp}

\begin{document}

\markboth{Wei Lu}
{Quark-gluon plasma and nucleons}

%
\catchline{}{}{}{}{}
%

\title{Quark-Gluon Plasma and Nucleons \`a la Laughlin}

\author{Wei Lu}

\address{Manhasset, NY 11030, USA\\
weiluphys@yahoo.com}

\maketitle

\begin{history}
\received{(Day Month Year)}
\revised{(Day Month Year)}
\end{history}

\begin{abstract}
Inspired by Laughlin's theory of the fractional quantum Hall effect, we explore the topological nature of the quark-gluon plasma and the nucleons in the context of the Clifford algebra. In our model, each quark is transformed into a composite particle via the simultaneous attachment of a spin monopole and an isospin monopole. This is induced by a novel kind of meson endowed with both spin and isospin degrees of freedom.  The interactions in the strongly coupled quark-gluon system are governed by the topological winding number of the monopoles, which is an odd integer to ensure that the overall wave function is antisymmetric. The states of the quark-gluon plasma and the nucleons are thus uniquely determined by the combination of the monopole winding number $m$ and the total quark number $N$. The radius squared of the quark-gluon plasma droplet is expected to be proportional to $mN$. We anticipate the observation of such proportionality in the heavy ion collision experiments.  
\end{abstract}

\keywords{Clifford algebra; geometric algebra; topological winding number; joint spin-isospin monopoles; quark-gluon plasma; quark confinement.}

\section{Introduction}
\label{sec:intro}

The color-charged particles, such as quarks and gluons, cannot be seen in isolation. The phenomenon of color confinement is posited as a natural property of quantum chromodynamics (QCD). However, the actual mechanism of confinement is still an open question. One line of thinking invokes topological solitons for the absolute confinement of quarks inside the hadronic bound states. It is conjectured by 't~Hooft~\cite{TH} and Mandelstam~\cite{MN} that the QCD vacuum is the electric-magnetic dual of a superconductor and monopoles will condensate to realize the confining phase. The topics of electric-magnetic duality and monopole condensation have also been explored by Seiberg and Witten~\cite{SW1, SW2} in the context of supersymmentic Yang-Mills theories.

Curiously, a new state of matter known as the quark-gluon plasma (QGP)~\cite{SH,BRS} has turned up many points of contact with monopoles as well~\cite{LS1,LS2,RS,LUO}. The QGP has been detected in the heavy ion collisions at the Relativistic Heavy Ion Collider (RHIC) at Brookhaven National Laboratory and the Large Hadron Collider (LHC) at CERN.  The QGP is arguably the most perfect liquid in nature, characterized by the remarkable feature of exceptionally low viscosity. Contrary to expectation, the QGP is not a weakly coupled plasma of color-charged particles. Rather, every quark and gluon is strongly coupled to its neighbors. The nonperturbative dynamics of the QGP poses a challenge to theoretical understanding. One approach of tackling this problem involves topologically nontrivial configurations known as the chromo-magnetic monopoles in the gluon sector~\cite{LS1,LS2}. This magnetic scenario suggests that the QGP is a magnetic plasma of non-condensed monopoles. Various studies~\cite{LS1,LS2,RS} indicate that the monopoles play an extremely important role in the QGP transport properties, which could be well described by the relativistic hydrodynamics.

The common thread of topological non-triviality running through the hadronic bound states and the QGP calls for a unified picture. We will henceforth use the terminology of `strongly coupled quark-gluon system' for both the nucleons and the QGP. The aim of the present paper is to depict the strongly coupled quark-gluon system by a general wave function. This is achieved by taking a page from Laughlin's celebrated theory~\cite{LG} of the fractional quantum Hall effect. The landmark discoveries of the quantum Hall effects have ushered in a new era of condensed matter physics. It has been realized that the quantized conductance of the integer quantum hall (IQH) effect~\cite{KL}  could be understood as a robust topological invariant known as the Chern or TKNN number~\cite{TKNN}, which led to the understanding of topological insulators and superconductors~\cite{HK,QZ}. Meanwhile, the fractional quantum Hall (FQH) effect~\cite{TS} has been elegantly explained by Laughlin’s wave function. The FQH states contain topological orders~\cite{WN,WEN,WEN2017} beyond Landau’s symmetry breaking paradigm. The topological quantum field theory (TQFT) description of the FQH effect has been subsequently developed based on the Chern-Simons gauge fields~\cite{ZH,BW1,BW2}.  

Motivated by an analogy with Laughlin's theory, we propound that each quark is simultaneously attached with a spin monopole and an isospin monopole in a strongly coupled quark-gluon system. The interactions in the system are controlled by the topological winding number of the monopoles, which is an odd integer to ensure that the total fermionic wave function is antisymmetric. Therefore, the states of the QGP and the nucleons are uniquely determined by the combination of the monopole winding number and the total quark number. Note that our work focuses on the spin and isospin monopoles, rather than the chromo-magnetic monopoles commonly considered with regard to the nucleons and the QGP. 

In this paper, we are going to resort to the Clifford algebra~\cite{HEST1,HEST2,PAV,Loun,DORA,WB} when we strive to extrapolate Laughlin's theory to the realm of the quark-gluon system. The Clifford algebra is also known as the geometric algebra.The Clifford algebra approach elegantly captures the non-Abelian nature of the spin and isospin operators and immensely facilitates the task of embedding the monopole configurations into the wave function. 

This paper is structured as follows: Section~\ref{sec:Laughlin} introduces Laughlin's theory of the FQH effect, and discusses its implications for the strongly coupled quark-gluon systems. In sect.~\ref{sec:MM}, we investigate a novel meson with both spin and isospin degrees of freedom, and study the joint spin-isospin monopoles formed by the meson field. In sect.~\ref{sec:SCQG}, we present the Clifford algebraic wave function, and make contact with the states of the QGP and the nucleons. In the last section we draw our conclusions. Throughout this paper, we adopt the units: $c = \hbar = 1$. 

%
%
%

\section{Lessons Learned from the Fractional Quantum Hall Effect}
\label{sec:Laughlin}

\subsection{Laughlin's Theory}
\label{sec:QHE}

Let us set the stage for our work by introducing Laughlin's wave function~\cite{LG} of the FQH liquid at filling fraction $\nu=1/m$,
\begin{align}
\label{eq:Lau}
\Psi_m( z_1\dots z_N)& = \prod_{i<j}^N (z_i- z_j)^m e^{-\sum_i^N  |z_i|^2/(4l_B^2)},
\end{align}
where $m$ is an odd integer, $N$ is the total electron number, $l_B$ is the magnetic length, and $z_i =x_i + iy_i$ is the location of the $i$th electron in terms of a complex number.  The number $m$ must be an odd integer, so that the wave function $\Psi_m$ is antisymmetric under the exchange of any two particles. Note that the case of $m=1$ reduces to the wave function of the IQH liquid with the fully filled Landau Level $\nu=1$.

In order to inspect the different facets of the wave function, we partition $\Psi_m$ into a unitary phase part $U_m( z_1\dots z_N)$ (with $U_m U_m^* =1$) and an amplitude part $e^{-\beta V_m( z_1\dots z_N)}$ (with $1/\beta = m$),
\begin{align}
\label{eq:Lau2}
\Psi_m( z_1\dots z_N) & = U_m( z_1\dots z_N)  e^{-\beta V_m( z_1\dots z_N) },
\end{align}
where
\begin{align}
\label{eq:U}
U_m( z_1\dots z_N) & = \prod_{i<j}^N \Big ( \frac{z_i- z_j}{|z_i- z_j|}\Big )^m,
\end{align}
and 
\begin{align}
\label{eq:V}
V_m( z_1\dots z_N) & =  -m^2\sum_{i<j}^N  \ln(|z_i- z_j|) + \frac{m}{4l_B^2}\sum_i^N  |z_i|^2.
\end{align}

Let us first examine the unitary $U_m( z_1\dots z_N)$, which can be cast into the form
\begin{align}
\label{eq:U}
U_m( z_1\dots z_N) & = e^{i\sum_{i<j}^N m\phi_{ij}},
\end{align}
where $\phi_{ij}$ is the azimuthal angle between the particles
\begin{align}
\label{eq:azimuthal}
\phi_{ij} & = \tan^{-1} \frac{y_i-y_j}{x_i-x_j}.
\end{align}

One can see that the phase $e^{im\phi_{ij}}$ is reminiscent of a vortex configuration. The unitary $U_m$ is describing the bound states of electrons attached with vortices of winding number $m$. In terms of the TQFT description of FQH effect~\cite{ZH}, these vortices correspond to Chern-Simons fluxes. Owing to the attachment of odd flux quanta (odd winding number $m$), the electrons are transmuted into bosonic composite particles. Note that the single-valuedness of the wave function demands that the vortex winding number $m$ must be an integer. The topological quantization of the vorticity underpins the exactness of the plateaus of the Hall resistance.

Now we turn to the amplitude part $e^{-\beta V_m( z_1\dots z_N)}$ of Laughlin's wave function. In an analogy with the classical plasma physics~\cite{LG}, the amplitude part is likened to the Boltzmann weight with inverse temperature  $1/\beta = m$. It is argued that $V_m( z_1\dots z_N)$ is the potential energy for a plasma of particles confined in two-dimensions, where each particle carries negative charge $q = -m$. The first term in eq.~\eqref{eq:V} is the Coulomb potential between two particles of charge $-m$, whereas the second term describes these particles moving in the potential of a positive-charged background. The constant charge density of the background is
\begin{align}
\label{eq:background}
\rho_B & = \frac{1}{2\pi l_B^2}.
\end{align}

In order to minimize the energy of the potential $V_m$, the plasma has to be charge-neutral on average. Considering that each particle carries a negative charge $-m$, the uniform compensating density of the particles should be
\begin{align}
\label{eq:density}
n & = \frac{\rho_B}{m} = \frac{1}{2\pi m l_B^2},
\end{align}
which corresponds to the electron density of the FQH liquid at the filling fraction $v= 1/m$.

\subsection{Implications for the Strongly Coupled Quark-Gluon Systems}
\label{sec:implications}

Laughlin's theory offers two valuable lessons for the investigation of the strongly coupled quark-gluon systems. The first takeaway is that there is a positive-charged background in the plasma analogy. It serves the purpose of neutralizing the negative charge of the electrons. In the case of a strongly coupled quark-gluon system, the counterpart of the background charge is the cloud of gluons which would hold the quarks together. In the flux-tube picture of QCD, the gluon field between a pair of quarks forms a string between them. Thus the commonly adopted phenomenological model of multi-quark potential is a sum of linear potentials between quark pairs. An alternative model, which is dubbed as the $Y$-law in the three-quark case~\cite{XA,IP,AN}, suggests that the potential grows linearly with the total length of a string network. These strings join at the Fermat point of the triangle formed by the three quarks, so that the total string length is minimized. In this paper, we take a similar view and propose that the quarks immersed in the sea of gluons experience an effective linear potential
\begin{align}
\label{eq:gluon-potential} 
V &= \sigma \sum_i^N  r_i,
\end{align}
where $\sigma$ is the string tension, N is the total number of quarks, and $r_i$ is the distance of the $i$th quark from the center of the strongly coupled quark-gluon system. The center is defined so that  the total string length of $\sum_i^N r_i$ is minimized. 

The second lesson we learned from Laughlin's theory is that there are vortex configurations buried in the waving function. A vortex in two dimensions is a topological soliton connected to the first homotopy group $\pi_1(S^1)$. For the FQH effect, one can neglect the spin degree of freedom, owing to Zeeman splitting in the presence of a magnetic field. Hence the target space of homotopy is $S^1$, i.e., the unitary phase part of the complex wave function. When it comes to the strongly coupled quark-gluon systems, the composition of the wave function and the target space of homotopy are much richer. With the expanded degrees of freedom, we are compelled to turn our attention to the topological solitons linked to the second homotopy group $\pi_2(S^2)$, which will be investigated in the next section.

\section{Mesons and Monopoles}
\label{sec:MM}

\subsection{Mesons with Both Spin and Isospin Degrees of Freedom}
\label{sec:meson}
As stated in the last section, the wave function of a strongly coupled quark-gluon system is more intricate than that of the FQH liquid. In addition to the basic ingredients such as quarks and gluons, QCD also involves composite states of mesons. This could have implications for the nature of the possible topological solitons.   

At low energies, QCD features $N_f=3$ flavors of light quarks. Nevertheless, we will focus on the two lightest flavors, i.e., up and down quarks, while omitting the strange quark contributions. Hence the terms such as flavor and isospin can be used interchangeably in the current paper. When the quark mass and the electroweak interactions are of minor concern compared with the strong force, the global symmetries of the two-flavor QCD are
\begin{align}
\label{eq:global1}
&  SU(2)_{iso_L} \times SU(2)_{iso_R} \times U(1)_B \times U(1)_A,
\end{align}
where $SU(2)_{iso_L}$ and $SU(2)_{iso_R}$ are two independent isospin symmetries acting on the left- and right-handed quark isospin doublets, respectively. While $U(1)_B$ is an exact symmetry associated with baryon number conservation, the axial $U(1)_A$ is not an exact symmetry since it is spoiled by the quantum anomaly. Due to the dynamical symmetry breaking via non-zero expectation values for the quark-antiquark bilinears, the chiral $SU(2)_{iso_L} \times SU(2)_{iso_R}$ symmetries are broken down to the vectorial 
\begin{align}
\label{eq:vector1}
SU(2)_{iso_V},
\end{align}
which rotates the left- and right-handed quarks synchronously. The pions, as the composite states of the quark-antiquark pair, are the resultant Nambu-Goldstone bosons~\cite{NJL}. 

In this work, we will generalize the above scenario. Our proposition is that the chiral symmetries could be expanded to subsume both spin and isospin degrees of freedom
\begin{align}
\label{eq:global1}
&  SU(2)_{spin_L} \times SU(2)_{spin_R} \times SU(2)_{iso_L} \times SU(2)_{iso_R},
\end{align}
where the extra $SU(2)_{spin_L}$ and $SU(2)_{spin_R}$ are two independent spin rotation symmetries acting on the left- and right-handed quarks, respectively~\footnote{The extra $SU(2)_{spin_L}$ and $SU(2)_{spin_R}$ could be part of the double cover of the left- and right-handed Lorentz symmetries $Spin(1,3)_{L}$ and $Spin(1,3)_{R}$ acting on the left- and right-handed quarks, respectively.}. The chiral symmetries are spontaneously broken down to the vectorial symmetries
\begin{align}
\label{eq:vector2}
SU(2)_{spin_V} \times SU(2)_{iso_V},
\end{align}
where $SU(2)_{spin_V} $rotates the spin of the left- and righ-handed quarks in the same way. Consequently, there is a novel kind of meson $\hat{\pi}$ endowed with both spin and isospin degrees of freedom, expressed as
\begin{align}
\label{eq:Pion}
\hat{\pi} = \sum_{i, j}\hat{\pi}^{i,j}\sigma^{spin}_i \otimes \sigma^{iso}_j,
\end{align}
where $i, j = 1, 2, 3$,  $\sigma^{spin}_i$ and $\sigma^{iso}_i$ are spin and isospin Pauli matrices, respectively. The $\hat{\pi}$ field could form topologically nontrivial configurations, which will be detailed in the next subsection.

Notice that the $\hat{\pi}$ meson is different from the vector mesons such as $\rho$ and $\omega$ mesons. The $\hat{\pi}$ meson is coupled to the spin current of quarks, whilst the vector mesons are coupled to the vector current of quarks. The vector mesons would not be considered in the current paper, albeit they could also play a vital role in the dynamics of QCD~\cite{SA}.

\subsection{The Joint Spin-Isospin Monopoles}
\label{sec:monopole}
Now we are are ready to examine the topological solitons formed by the meson field $\hat{\pi}(\vec{r})$. The joint spin-isospin monopoles can be written as
\begin{align}
\label{eq:monopole}
(\hat{n}_m(\vec{r}) \cdot \vec{\sigma}^{spin}) \otimes (\hat{n}_m(\vec{r}) \cdot \vec{\sigma}^{iso}),
\end{align}
where $\hat{n}_m(\vec{r})$ is a three-dimensional unit vector with components
\begin{subequations}\label{eq:unit}
\begin{align}
\hat{n}_m^x(\vec{r}) &= \sin\theta\cos(m\phi), \\
\hat{n}_m^y(\vec{r}) &= \sin\theta\sin(m\phi), \\
\hat{n}_m^z(\vec{r}) &= \cos\theta.
\end{align}
\end{subequations}
Here $\theta \in [0, \pi]$ is the polar angle of the three-dimensional position vector $\vec{r}$, and $\phi \in [0, 2\pi)$ is the azimuthal angle of $\vec{r}$. The monopole is characterized by the topological winding number $m$, which is required to be an integer in order for $\hat{n}_m(\vec{r})$ to be single-valued and smooth everywhere. The spherically symmetric case of $m=1$ shares the same hedgehog ansatz of the non-Abelian magnetic monopoles propounded by ’t~Hooft~\cite{THO} and Polyakov~\cite{POL}. The nontrivial spin (isospin) configuration corresponds to the map from the positional sphere $S^2$ into the spin (isospin) unit vector sphere $S^2$. As such, the topologically distinct maps belong to the second Homotopy group $\pi_2(S^2)$. 

When $m$ is an odd integer, it can be verified that the joint spin-isospin monopoles are invariant under the reversion of the position vector $\vec{r} \rightarrow -\vec{r}$. This property is of paramount importance when we set out to attach the joint spin-isospin monopoles to the quarks to construct a suitable wave function for the strongly coupled quark-gluon systems. Note that the spin monopole or the isospin monopole individually changes sign under the spacial reversion. 

It is worth mentioning that there is another category of solitons introduced by Skyrme~\cite{SKY} with a view toward understanding the nucleons. Skyrme has demonstrated that the nonlinear field theory of pions could have topological nontrivial solutions known as Skyrmions. It has been suggested that proton and neutron might actually be Skyrmions. The Skyrmion theory has been perceived as a low energy effective theory of QCD in the limit of a large number of quark colors~\cite{WIT}. A Skyrmion entails a map from $\mathbb{R}^3 \cup \{\infty\}$ into $S^3$, the group manifold of $SU(2)$. It follows that Skymions fall under the umbrella of the third homotopy group $\pi_3(S^3)$, differing from the second homotopy group $\pi_2(S^2)$ of the spin-isospin monopoles discussed above.

\section{The Wave Function of the Strongly Coupled Quark-Gluon Systems} 
\label{sec:SCQG}
\subsection{The Clifford algebra approach}
We are finally poised to undertake the task of formulating the wave function of the strongly coupled quark-gluon systems. And yet we are confronted with a hindrance: a fermion such as a quark is represented by a column spinor, whereas the spin-isospin monopoles are tied to the Pauli matrices. Due to the dichotomy between particle states and matrix operators acting on them in the conventional formalism of quantum field theory (QFT), it is rather cumbersome when we endeavor to weave the spin-isospin monopole configurations into the wave function. 

In view of the above, we are going to enlist the aid of  the Clifford algebra approach, whereby both the algebraic spinor states and Dirac's gamma operators can be expressed in the same algebraic space. The Clifford algebra, also known as the geometric algebra, is a potent mathematical apparatus that finds extensive applications in the arena of physics~\cite{HEST1,HEST2,PAV,Loun,DORA,WB}. The specific Clifford algebra $Cl(0,6)$ can be exploited to establish a unified theory of the standard model and gravity~\cite{WL1, WL2}. The standard model fermions are represented by the algebraic spinors of $Cl(0,6)$, while gravity is governed by the gauge theory of the Lorentz symmetry. It turns out that there is more to the Dirac operator $\gamma_0$ than meets the eye: $\gamma_0$ is surreptitiously a composite operator corresponding to the $Cl(0,6)$ tri-vector $\gamma_0 = \Gamma_1\Gamma_2\Gamma_3$. This revelation affords the possibility of fusing the gravitational and standard model interactions into a cohesive whole.  The Clifford algebra $Cl(0,6)$ can naturally accommodate the beyond-standard-model gauge symmetries $SU(3)_c \times SU(2)_L \times U(1)_R \times U(1)_{B-L}$ as well as global chiral symmetries in the extended top condensation model~\cite{WL3}, which provides considerable insight into the standard model fermion mass hierarchies and flavor mixing by taking advantage of four-fermion condensations and a troika of Higgs doublets. The Clifford algebra can also be brought to bear on modifying the gauge gravity theory and explaining the accelerated expansion of the universe via introducing a characteristic Hubble scale $h_0$ as a new fundamental physical constant~\cite{WL4, WL5}. The recent development of the Clifford functional integral formalism~\cite{WL5} illustrates that the Clifford algebra approach can further penetrate into the previously inaccessible QFT domains such as field quantization (second quantization), the derivation of the functional-differential Schwinger-Dyson equation, and the calculation of the beyond-tree-level quantum-loop corrections to propagators.

Before proceeding to choose the proper Clifford algebra to be used in this paper, let us recall that we utilize the effective linear potential~\eqref{eq:gluon-potential} to depict the quark-gluon interactions. We also make the assumption that quark-quark interactions can be effectively described by the exchange of the spin-isospin $\hat{\pi}$ mesons.  Hence our model would involve the spin and isospin degrees of freedom, whereas the color degree of freedom would be deemed as an implicit variable.  It should be highlighted that the color degree of freedom manifests itself in the form of the stringy potential and the meson-mediated interactions, even though it is not explicitly present in the wave function. 

While the Clifford algebra $Cl(0,6)$ comprises all the standard model gauge couplings such as the electroweak and QCD strong interactions, in the current work we are going to focus on the simpler and familiar $Cl(1,3)$, which also goes under the name of Dirac algebra or spacetime algebra. Given that an algebraic spinor of $Cl(1,3)$ can entertain both spin and isospin degrees of freedom~\cite{HEST3, WL5},  the Clifford algebra $Cl(1,3)$ is thus a preferable choice for an effective representation of a strongly coupled quark-gluon system. 

The Clifford algebra $Cl(1,3)$ is defined by the vector basis $\{\gamma_{\mu}; \mu= 0, 1, 2, 3\}$ satisfying
\begin{align}
\label{eq:gamma}
&\gamma_{\mu}\gamma_{\nu}+\gamma_{\nu}\gamma_{\mu} = 2\eta_{\mu\nu},
\end{align}
where $\eta_{\mu\nu} = diag(1, -1, -1, -1)$.

The wave function $\Psi$ of an algebraic spinor is a linear combination of all $2^{4}=16$ basis elements of the Clifford algebra $Cl(1,3)$,
\begin{align}
\label{eq:psi}
\Psi =&\psi_0 + \sum_\mu \psi_1^{\mu}\gamma_\mu + \sum_{\mu<\nu}\psi_2^{\mu\nu}\gamma_{\mu}\gamma_{\nu} + \sum_\mu \psi_3^{\mu}i\gamma_{\mu} + \psi_4 i,
\end{align}
where  the unit pseudoscalar 
\begin{align}
i = \gamma_0\gamma_{1}\gamma_{2}\gamma_{3},
\end{align}
squares to $-1$, anticommutes with Clifford-odd elements, and commutes with Clifford-even elements. The $16$ linear combination coefficients such as $\psi_0$, $\psi_1^{\mu}$, etc. are real numbers~\footnote{In this paper, we are interested in $\Psi$ as a wave function. On the other hand, when $\Psi$ is treated as a fermion field in terms of the Clifford functional integral formalism~\cite{WL5}, the linear combination coefficients such as $\psi_0$, $\psi_1^{\mu}$, etc. are real Grassmann numbers, rather than real numbers.}. Note that the wave function~\eqref{eq:psi} should not be confused with a bispinor, which is effectively bosonic and can also be expanded in terms of the 16 elements of $Cl(1,3)$. The interested readers shall refer to refs.~\cite{HEST1,HEST2,PAV,Loun,DORA} and especially sect. 4.1 in ref.~\cite{PAV2} for detailed expositions on the mapping between a real algebraic spinor and a conventional complex column spinor. 

The wave function of an algebraic spinor of $Cl(1,3)$ can be identified with an isospin doublet~\cite{HEST3, WL5}
\begin{align}
&\Psi = \Psi_{+} + \Psi_{-},
\end{align}
where $\Psi_{+}$ and $ \Psi_{-}$ correspond to the isospin up and down projections of $\Psi$, respectively,
\begin{align}
\Psi_{\pm} &= \Psi P_{\pm},
\end{align}
with the isospin projection operators defined as
\begin{align}
\label{eq:projection}
P_{\pm} &= \frac{1 \pm {\gamma_3}{\gamma_0}}{2}.
\end{align}

Under the spin-isospin $SU(2)_{spin_V} \times SU(2)_{iso_V}$ rotations \eqref{eq:vector2}, the algebraic spinor $\Psi$ transforms as
\begin{align}
\label{eq:transform}
\Psi \rightarrow e^{\frac{1}{2}(\varepsilon_x \gamma_2\gamma_3 + \varepsilon_y \gamma_3\gamma_1 + \varepsilon_z \gamma_1\gamma_2)}\Psi e^{\frac{1}{2}(\vartheta_x \gamma_2\gamma_3 + \vartheta_y \gamma_3\gamma_1 + \vartheta_z \gamma_1\gamma_2)},
\end{align}
where $\vec{\varepsilon}$ and $\vec{\vartheta}$ are spin and isospin rotation parameters, respectively. Of particular note is that the spin and isospin rotation operators share the same algebraic formulae
\begin{align}
\label{eq:Pauli}
\gamma_2\gamma_3, \quad \gamma_3\gamma_1, \quad \gamma_1\gamma_2.
\end{align}
The differentiation between spin and isospin operators rests on the fact that the former shall be applied to the left-hand side of an algebraic spinor, whilst the latter shall be applied to the right-hand side. This is the underlying reason that an algebraic spinor of $Cl(1,3)$ can allow for both spin and isospin degrees of freedom, which is one of the advantages of the Clifford algebra approach compared with the conventional column spinor formalism. Note that historically various unification models~\cite{TRAY, WL1, WL2, WL3} have leveraged the liberty of applying disparate transformations on the left- and right-hand sides of an algebraic spinor.

\subsection{The phase-decorated part of the wave function}
\label{subsec:phase}

Equipped with the insight gained from Laughlin's theory of the FQH effect, we cast the algebraic wave function (apart from an overall normalization factor) of a strongly coupled quark-gluon system as a union of a phase-decorated part $\hat{\Psi}_{m, N}(\vec{r}_c;\vec{r}_1\dots \vec{r}_N)$  and an amplitude-modulation part $e^{- \beta V_{m, N}(\vec{r}_c;\vec{r}_1\dots \vec{r}_N)}$, 
\begin{align}
\label{eq:psim}
\Psi_{m, N}(\vec{r}_c;\vec{r}_1\dots \vec{r}_N) & = \hat{\Psi}_{m, N}(\vec{r}_c;\vec{r}_1\dots \vec{r}_N)  e^{-\beta V_{m, N}(\vec{r}_c;\vec{r}_1\dots \vec{r}_N) },
\end{align}
where $1/\beta = m\Lambda$, $\Lambda$ is a characteristic QCD energy scale of the order of the Hagedorn temperature, $m$ is an odd integer, $N$ is the total number of quarks, $\vec{r}_i$ is the position of the $i$th quark, and $\vec{r}_c$ is the position of the center of the  system. The center is defined so that $\vec{r}_c$ minimizes the total string length of $\sum_i^N|\vec{r}_i-\vec{r}_c|$. Note that $\Psi_{m, N}(\vec{r}_c;\vec{r}_1\dots \vec{r}_N)$ and $\hat{\Psi}_{m, N}(\vec{r}_c;\vec{r}_1\dots \vec{r}_N)$ are valued in the algebraic spinor space of $Cl(1,3)$ as in eq.~\eqref{eq:psi}, while $e^{- \beta V_{m, N}(\vec{r}_c;\vec{r}_1\dots \vec{r}_N)}$ and $V_{m, N}(\vec{r}_c;\vec{r}_1\dots \vec{r}_N)$ are valued in the real number space. 

Before delving into the details, we would like to make a few observations. First of all, the wave function is presumably describing a system of total spin $\frac{1}{2}$ and total isospin $\frac{1}{2}$. Furthermore, as noted earlier we only take into consideration the two lightest flavors of quarks, i.e., up and down quarks. Therefore, the wave function could cover the nucleons such as protons and neutrons, while the baryons with higher spin or non-zero strangeness are not considered in this work. Similarly, the QGP states studied in the current paper is also confined to the case of total spin $\frac{1}{2}$ and total isospin $\frac{1}{2}$.

Secondly, a self-contained strongly coupled quark-gluon system has a total color charge of zero. Hence the number of quarks in the system is a multiple of three, i.e., $N=3B$, where  $B$ is the total baryon number. It shall be underscored that $N$ is the number of valence quarks, rather than the indefinite number of virtual sea quarks and antiquarks. Because of the total spin and isospin constraints mentioned above, we assume that $B$ is an odd integer so that $N = 3, 9, 15, \cdots$. In other words, we assume that there are odd number of quarks in the system. 

And lastly, a color singlet is implicitly color-antisymmetric, albeit the color degree of freedom is not explicitly present in the wave function. Consequently the wave function~\eqref{eq:psim}, which involves the spin, isospin, and position degrees of freedom, shall be symmetric under the exchange of two quarks. This is to ensure that the wave function as a whole, accounting for both explicit and implicit degrees of freedom, is antisymmetric. 

With that, let us specify the phase-decorated part of the wave function
\begin{align}
\label{eq:phase}
\hat{\Psi}_{m, N}(\vec{r}_c;\vec{r}_1\dots \vec{r}_N)  & = \sum_{p} U_{m, N, p}(\vec{r}_1\dots \vec{r}_N)\hat{\Psi}_{0}(\vec{r}_c)U^{\dagger}_{m, N, p}(\vec{r}_1\dots \vec{r}_N),
\end{align}
where
\begin{align}
\label{eq:Um}
U_{m, N, p}( \vec{r}_1\dots \vec{r}_N) = P\bigg{[}\prod_{i<j}^N \Big{(}\hat{n}_m^x(\vec{r}_i-\vec{r}_j)\gamma_2\gamma_3 + \hat{n}_m^y(\vec{r}_i-\vec{r}_j) \gamma_3\gamma_1 + \hat{n}_m^z(\vec{r}_i-\vec{r}_j) \gamma_1\gamma_2\Big{)}\bigg{]}. 
\end{align}
Note that $\hat{\Psi}_{0}(\vec{r}_c)$ is the bare algebraic wave function of the strongly coupled quark-gluon system. The bare wave function $\hat{\Psi}_{0}(\vec{r}_c)$ represents the system as a whole as if there were no internal structure such as the constituent quarks. The bare wave function $\hat{\Psi}_{0}(\vec{r}_c)$ is to be embroidered with the phases $U_{m, N, p}(\vec{r}_1\dots \vec{r}_N)$, $U^{\dagger}_{m, N, p}(\vec{r}_1\dots \vec{r}_N)$, and the amplitude moderation $e^{- \hat{V}_{m, N}}$ to constitute a comprehensive representation of the quark-gluon ensemble. 

The unit vector $\hat{n}_m(\vec{r}_i-\vec{r}_j)$ appearing in $U_{m, N, p}(\vec{r}_1\dots \vec{r}_N)$ is defined by eq.~\eqref{eq:unit}, which is characterized by the topological winding number $m$ of the second homotopy group $\pi_2(S^2)$. It corresponds to the monopole configurations induced by the spin-isospin $\hat{\pi}$ mesons as discussed in the previous section. The $P[ \cdots ]$ notation on the right-hand side of eq.~\eqref{eq:Um} stands for a given permutation of the order in the product $\prod_{i<j}^N$ of the $(i<j)$ pairs. A specific permutation is labeled by the subscript $p$ in $U_{m, N, p}(\vec{r}_1\dots \vec{r}_N)$. Owing to the non-Abelian nature of the spin-isospin operators~\eqref{eq:Pauli}, a different order in the multiplication of $(i<j)$ pairs gives rise to a different product. On the other hand, the multiplication order of the Abelian phases does not matter for the wave function of the FQH states. Note that the summation $\sum_{p}$ in eq.~\eqref{eq:phase} ensures that all the different permutations are included. 

The Hermitian conjugate of $U^{\dagger}_{m, N, p}(\vec{r}_1\dots \vec{r}_N)$ takes the form
\begin{equation}
\label{eq:Her}
U^{\dagger}_{m, N, p}(\vec{r}_1\dots \vec{r}_N) = \gamma_0\tilde{U}_{m, N, p}(\vec{r}_1\dots \vec{r}_N)\gamma_0,
\end{equation}
where the reversion of $U_{m, N, p}(\vec{r}_1\dots \vec{r}_N)$, denoted $\tilde{U}_{m, N, p}(\vec{r}_1\dots \vec{r}_N)$, reverses the order in any product of Clifford vectors. It can be  shown that $U_{m, N, p}$ is unitary 
\begin{equation}
\label{eq:unitary}
U_{m, N, p}(\vec{r}_1\dots \vec{r}_N)U^{\dagger}_{m, N, p}(\vec{r}_1\dots \vec{r}_N) = 1. 
\end{equation} 

The unitary $U_{m, N, p}(\vec{r}_1\dots \vec{r}_N)$ plays a similar role as the unitary phase part of Laughlin's wave function of the FQH effect. Analogous to the FQH case where electrons are bundled with vortices, the phase-decorated part of the wave function $\hat{\Psi}_{m, N}(\vec{r}_c;\vec{r}_1\dots \vec{r}_N)$ represents the bound states of quarks attached with monopoles. As we learned from the spin-isospin transformations~\eqref{eq:transform},  the spin operators are applied to the left-hand side of an algebraic spinor, whereas the isospin operators are applied to the right-hand side. It follows that the phase factors $U_{m, N, p}(\vec{r}_1\dots \vec{r}_N)$ and $U^{\dagger}_{m, N, p}(\vec{r}_1\dots \vec{r}_N)$ simultaneously attach each quark with a spin monopole and an isospin monopole characterized by the topological winding number $m$. 

When $m$ is an odd integer, it has been demonstrated earlier that the joint spin-isospin monopoles~\eqref{eq:monopole} are invariant under the reversion of the position vector $\vec{r} \rightarrow -\vec{r}$, whereas the spin monopole or the isospin monopole individually changes sign upon spacial reversion. Aided by this property, it can be verified that the wave function $\hat{\Psi}_{m, N}(\vec{r}_c;\vec{r}_1\dots \vec{r}_N)$ is symmetric if we switch places of any pair of quarks. And it can be checked that the amplitude-modulation part of the wave function presented in the next subsection is also symmetric. Since we know that a color singlet is implicitly color-antisymmetric, the wave function that takes together all degrees of freedom is thereby antisymmetric. This is the desired feature of a fermionic many-particle wave function. Note that the symmetric summation of all the possible permutations in eq.~\eqref{eq:phase} is part and parcel of the overall antisymmetrization scheme. It has to be accentuated that the composite particle of a quark simultaneously attached with a spin monopole and an isospin monopole is still a fermion. This is in contrast to Laughlin's wave function of the FQH effect, where the electrons are transmuted into bosonic composite particles due to the attachment of odd flux quanta.

\subsection{The amplitude-modulation part of the wave function}
\label{subsec:amplitude}
In parallel with Laughlin's plasma analogy~\cite{LG}, $V_{m, N}$ in the amplitude-modulation part of the wave function~\eqref{eq:psim} can be written as the potential energy for a plasma of particles moving in three-dimensions
\begin{align}
\label{eq:Vm}
V_{m, N}(\vec{r}_c;\vec{r}_1\dots \vec{r}_N) & =  m^2 \sum_{i<j}^N  \frac{1}{|\vec{r}_i-  \vec{r}_j|} + \frac{m}{l_\sigma^2} \sum_i^N  |\vec{r}_i-  \vec{r}_c|,
\end{align}
where the second term corresponds to the effective potential~\eqref{eq:gluon-potential} in the flux-tube picture of the gluon field, and $1/l_\sigma^2$ reparameterizes the string tension for the linear potential.  

The first term in eq.~\eqref{eq:Vm} is interpreted as the three-dimensional ``Coulomb'' potential between two particles of negative charge $q=-m$, whereas the second term can be construed as describing these particles moving in the potential of a positive-charged background. Given that the linear potential $\phi(\vec{r}) = \frac{m}{l_\sigma^2}  |\vec{r}-  \vec{r}_c|$ in the second term of eq.~\eqref{eq:Vm} obeys the Poisson equation
\begin{align}
\label{eq:Poisson}
-\nabla^2\phi(\vec{r}) & = 4\pi q \rho(\vec{r}),
\end{align}
the charge density $\rho(\vec{r})$ of the background can be calculated as
\begin{align}
\label{eq:rho}
\rho(\vec{r}) & = \frac{1}{2\pi l_\sigma^2} \frac{1}{|\vec{r}-\vec{r}_c|}.
\end{align}

In order to minimize the energy of the potential $V_{m, N}$, the plasma has to be charge-neutral on average. Considering that each particle carries a negative charge $-m$, the compensating density of the particles should be
\begin{align}
\label{eq:densityr}
n(\vec{r}) & = \frac{\rho(\vec{r})}{m} = \frac{1}{2\pi m l_\sigma^2} \frac{1}{|\vec{r}-\vec{r}_c|}.
\end{align}

Some clarification is in order at this point. The particles mentioned above are composite particles as delineated by the phase-decorated part of the wave function~\eqref{eq:phase}. By construction, each particle is composed of a quark, a spin monopole, and an isospin monopole.  Because of the monopole attachment, the original strong couplings between quarks are reincarnated as the ``Coulomb'' interactions embedded in the amplitude-modulation part of the wave function. The charge $q=-m$ of the particles is predetermined by the topological winding number $m$ of the monopoles.   The charge plays a central role in the particle-particle as well as particle-background interactions, which in turn would dictate the density~\eqref{eq:densityr} of the particles. It is worth mentioning that the positive and negative charges alluded above shall not be conflated with either color or electric charges. 

Since the composite particles and the quarks are in one-to-one correspondence, the quark density is the same as the particle density.  In contrast to the uniform density~\eqref{eq:density} of the electrons in the FQH liquid, the quark density~\eqref{eq:densityr} is non-uniform. The density is higher toward the center $\vec{r}_c$ of the system.  Note that the discussion here concerns the density of the valence quarks, while the seething sea of virtual quarks and antiquarks could assume a different density profile. 

Given the density function and the total number $N$ of quarks, the radius $R$ of the strongly coupled quark-gluon system should satisfy
\begin{align}
\label{eq:volume}
\int_{|\vec{r}-\vec{r}_c|\le R}n(\vec{r})dr^3 = N,
\end{align}
which can be employed to calculate the radius as
\begin{align}
\label{eq:radius}
R = \sqrt{mN} l_\sigma.
\end{align}
It follows that the average quark density reads
\begin{align}
\label{eq:densityR}
 \bar{n}& = \frac{1}{m^{\frac{3}{2}} N^{\frac{1}{2}} }\frac{1}{\frac{4}{3}\pi l_\sigma^3}=\frac{3}{2}n(\vec{r})|_{|\vec{r}-\vec{r}_c| = R},
\end{align}
where $n(\vec{r})|_{|\vec{r}-\vec{r}_c| = R}$ is the density of quarks at the surface of the system. Evidently $l_\sigma$ serves as the characteristic length scale, whereas the characteristic energy scale of the system is $\Lambda$ which appears in the amplitude-modulation part of the wave function~\eqref{eq:psim} in the form of $1/\beta = m\Lambda$.

\subsection{The nucleons and the quark-gluon plasma}
\label{subsec:nucleons}
After fleshing out the details of the waving function~\eqref{eq:psim}, we are now well positioned to assign the wave function to the particular instances of the strong-correlated quark-gluon systems. Let us start with 
\begin{align}
\label{eq:m1n3}
\Psi_{m=1, N=3}(\vec{r}_c;\vec{r}_1, \vec{r}_2, \vec{r}_3),
\end{align}
which involves the minimal number of quarks $N=3$ in a color-neutral system. Here $\vec{r}_c$ is the Fermat point of the triangle formed by $\vec{r}_1$, $\vec{r}_2$, and $\vec{r}_3$. We propound that the wave function can be identified with the proton-neutron isospin doublet. In other words, the wave functions of a proton and a neutron are, respectively,
\begin{subequations}\label{eq:nucleons}
\begin{align}
\label{eq:proton}
\Psi_{proton}=\Psi_{m=1, N=3}(\vec{r}_{proton};\vec{r}_1, \vec{r}_2, \vec{r}_3)P_{+},
\end{align}
and
\begin{align}
\label{eq:neutron}
\Psi_{neutron}=\Psi_{m=1, N=3}(\vec{r}_{neutron};\vec{r}_1, \vec{r}_2, \vec{r}_3)P_{-},
\end{align}
\end{subequations}
where $P_{\pm}$ are the isospin projectors~\eqref{eq:projection}.

The wave functions suggest that the quarks in a nucleon are simultaneously bundled with the spin and isospin monopoles of winding number $m=1$. As discussed earlier, the topological number $m$ of a joint spin-isospin monopole has to be an odd integer, so that the overall waving function including the color degree of freedom is antisymmetric upon switching places of any pair of quarks. Hence $|m| = 1$ is the minimal absolute value it can take. 

For the other allowable combinations of $m$ and $N$, our thesis is that the wave function $\Psi_{m, N}(\vec{r}_c;\vec{r}_1\dots \vec{r}_N)$ as defined in eq.~\eqref{eq:psim} can be associated with the state of the QGP detected in the heavy ion collisions at the RHIC and the LHC~\cite{SH, BRS}. Note that the proposed algebraic waving function describes a system with  spin $\frac{1}{2}$ and isospin $\frac{1}{2}$, therefore it only applies to the QGP states with such total spin and isospin numbers. In the QGP, every quark and gluon is strongly coupled to its neighbors. By virtue of the spin-isospin monopole attachment to the quarks, the strong coupling is transmuted into the ``Coulomb'' interactions encoded in the wave function. 

Since the wave function $\Psi_{m=1, N=3}$ is describing the nucleons, it is tempting to speculate that the $\Psi_{m=1, N>3}$ wave functions might correspond to nuclides. However, it is not the case. A nuclide (other than the lightest $^1$H) is composed of a cluster of protons and neutrons, with each portrayed by the above wave function $\Psi_{proton}$ or $\Psi_{proton}$  individually.  As such, the quarks within a given nucleon are indeed strongly correlated with each other. On the other hand, the inter-nucleon interactions, such as the electrostatic repulsion and the residual interactions mediated by the escaping mesons, are much weaker. This is dissimilar to the case of the wave function $\Psi_{m=1, N>3}$, where all quarks in the plasma are inextricably entangled with each other. Therefore, the wave function $\Psi_{m=1, N>3}$ depicts the state of the QGP, instead of nuclides.

For the QGP associated with the wave function $\Psi_{m>1, N}$, the quark density in the plasma is lower (see eq.~\eqref{eq:densityr}) than the QGP with $m=1$. The reason is that the  charge $q=-m$ of the particles in the plasma is determined by the topological winding number $m$.  A larger $m$ is translated into a stronger repulsive ``Coulomb'' interaction which pushes the particles further away from each other. For a given quark number $N$, the QGP with a larger $m$ would have a larger volume. The radius of the QGP droplet is proportional to $\sqrt{mN}$ as indicated by eq.~\eqref{eq:radius}. We anticipate the experimental verification of such proportionality in the future probes of the QGP. Considering that the monopole configuration~\eqref{eq:unit} for $m>1$ is not spherically symmetric, we assume that the $z$ axis of the wave function is aligned with the impact direction of the heavy ion collisions.

As a last note, we shall mention that a larger $m$ or a larger $N$ implies that more particles are  distributed further away from the center and subject to higher string potential~\eqref{eq:gluon-potential}. As such, the QGP corresponds to a higher energy (or higher temperature) metastable state compared with the stable states such as the $\Psi_{m=1, N=3}$ nucleons. Therefore the QGP states emerging from the heavy ion collisions would eventually decay to the more stable hadronic matter as observed at the RHIC and the LHC~\cite{SH, BRS}. 

\section{Conclusions}
\label{sec:concl}
The new state of matter known as the quark-gluon plasma (QGP) has received much attention due to its singular properties. From the new vantage point of knowledge gained from the QGP, we may glean a helpful clue or two for deciphering the mechanism of color confinement. The goal of the current paper is to provide a unified framework for both the QGP and the confining states such as protons and neutrons. 

Inspired by Laughlin's theory of the fractional quantum Hall (FQH) effect, we postulate a Clifford algebraic general wave function for the QGP as well as the nucleons. In our proposal, each quark is transmuted into a composite particle via the simultaneous attachment of a spin monopole and an isospin monopole. This is facilitated by a novel type of meson endowed with both spin and isospin degrees of freedom. This particular meson stems from the spontaneous symmetry breaking of the chiral spin and chiral isospin degrees of freedom $SU(2)_{spin_L} \times SU(2)_{spin_R} \times SU(2)_{iso_L} \times SU(2)_{iso_R}$. 

The interactions in the strongly coupled quark-gluon system are controlled by the monopole winding number $m$, which characterizes the topological solitons of the second homotopy group $\pi_2(S^2)$.  The number $m$ is an odd integer to ensure that the total fermionic wave function is antisymmetric upon switching places of any pair of quarks. The states of the QGP and the nucleons are thereby uniquely determined by the combination of the topological number $m$ and the total quark number $N$. The wave function $\Psi_{m=1, N=3}$ is identified with the proton-neutron isospin doublet. It corresponds to the minimal absolute value of $|m|=1$ and the minimal number of quarks $N=3$ in a color-neutral system. For the other admissible combinations of $m$ and $N$, the wave function $\Psi_{m, N}$ can be associated with the QGP detected at the RHIC and the LHC. The radius  $R$ of the QGP is expected to be proportional to $\sqrt{mN}$. We anticipate the verification of such proportionality in the heavy ion collision experiments. 

Our work is originated from an analogy between the FQH effect and the strongly coupled quark-gluon system. One of the noteworthy features of the FQH state is the fractional excitations that carry fractional charges~\cite{LG} and fractional statistics~\cite{HA, ASW}. These are also referred to as the topological excitations~\cite{WEN2017},  given that these excitations cannot be created by the local operators. It stands to reason to wonder whether there are analogous topological excitations in the strongly coupled quark-gluon systems. We leave this interesting topic to future studies. 

Additionally, we hope that our approach could serve as a steppingstone to the ultimate theory of neutron stars. The LIGO-Virgo collaboration's observation of the gravitational waves from a binary neutron star inspiral~\cite{LV} has rejuvenated the interest in the neutron stars. The neutron star phenomenology has been a proving ground for myriads of nuclear theory models and modified gravity models. We propound that the joint spin-isospin monopoles contemplated in this paper might play an eminent role in the core of a neutron star. As such, the phase sector of the wave function of the neutron star core could resemble the one delineated in the current work. In light of the fact that the mass of a neutron star is extremely concentrated, the enormous gravitational field shall be factored into consideration in conjunction with the QCD interactions. The collective effect could have consequential implications for the amplitude part of the wave function.


\begin{thebibliography}{0}
\bibitem{TH} G. 't Hooft, Topology of the gauge condition and new confinement phases in non-Abelian gauge theories, {\it Nucl. Phys. B }{\bf 190}, 455 (1981).

\bibitem{MN} S. Mandelstam, Vortices and quark confinement in non-Abelian gauge theories, {\it Phys. Rep. }{\bf 23}, 245 (1976). 

\bibitem{SW1} N. Seiberg and E. Witten, Electric-magnetic duality, monopole condensation, and confinement in N= 2 supersymmetric Yang-Mills theory, {\it Nucl. Phys. B }{\bf 426}, 19 (1994).

\bibitem{SW2} N. Seiberg and E. Witten, Monopoles, duality and chiral symmetry breaking in N= 2 supersymmetric QCD, {\it Nucl. Phys. B }{\bf 431}, 484 (1994).

\bibitem{SH} E. Shuryak, Strongly coupled quark-gluon plasma in heavy ion collisions, {\it Rev. Mod. Phys. }{\bf 89}, 035001 (2017).

\bibitem{BRS} W. Busza, K. Rajagopal and W. van der Schee, Heavy ion collisions: the big picture and the big questions, {\it Annu. Rev. Nucl. Part. Sci. }{\bf 68}, 339 (2018).

\bibitem{LS1} J. Liao and E. Shuryak, Strongly coupled plasma with electric and magnetic charges, {\it Phys. Rev. C }{\bf 75}, 054907 (2007).

\bibitem{LS2} J. Liao and E. Shuryak, The magnetic component of quark-gluon plasma is also a liquid, {\it Phys. Rev. Lett. }{\bf 101}, 162302 (2008).

\bibitem{RS} C. Ratti and E. Shuryak, Role of monopoles in a gluon plasma, {\it Phys. Rev. D }{\bf 80}, 034004 (2009).

\bibitem{LUO} M. J. Luo, Quark–gluon plasma and topological quantum field theory, {\it Mod. Phys. Lett. A }{\bf 32}, 1750056 (2017).

\bibitem{LG} R. B. Laughlin, Anomalous quantum Hall effect: an incompressible quantum fluid with fractionally charged excitations, {\it Phys. Rev. Lett. }{\bf 50}, 1395 (1983).

\bibitem{KL} K. Klitzing, G. Dorda and M. Pepper, New method for high-accuracy determination of the fine-structure constant based on quantized Hall resistance, {\it Phys. Rev. Lett. }{\bf 45}, 494 (1980).

\bibitem{TKNN} D. J.  Thouless {\it et al.}, Quantized Hall conductance in a two-dimensional periodic potential, {\it Phys. Rev. Lett. }{\bf 49}, 405 (1982).

\bibitem{HK} M. Z. Hasan, C. L. Kane, Colloquium: Topological insulators, {\it Rev. Mod. Phys. }{\bf 82}, 3045 (2010).

\bibitem{QZ} X.-L. Qi and S.-C. Zhang, Topological insulators and superconductors, {\it Rev. Mod. Phys. }{\bf 83}, 1057 (2011).

\bibitem{TS} D. C.  Tsui, H. L. Stormer and A. C. Gossard, Two-dimensional magnetotransport in the extreme quantum limit, {\it Phys. Rev. Lett. }{\bf 48}, 1559 (1982).

\bibitem{WN} X.-G. Wen and Q. Niu, Ground-state degeneracy of the fractional quantum Hall states in the presence of a random potential and on high-genus Riemann surfaces, {\it Phys. Rev. B }{\bf 41}, 9377 (1990).

\bibitem{WEN} X.-G. Wen,  {\it Quantum Field Theory of Many-body Systems: From the Origin of Sound to an Origin of Light and Electrons}, (Oxford University Press, Oxford, 2004).

\bibitem{WEN2017} X.-G. Wen, Colloquium: Zoo of quantum-topological phases of matter, {\it Rev. Mod. Phys. }{\bf 89}, 041004 (2017).

\bibitem{ZH} S.-C. Zhang, T. H. Hansson and S. Kivelson, Effective-field-theory model for the fractional quantum Hall effect, {\it Phys. Rev. Lett. }{\bf 62}, 82 (1989).

\bibitem{BW1} B. Blok and X. -G. Wen, Effective theories of the fractional quantum Hall effect at generic filling fractions, {\it Phys. Rev. B }{\bf 42}, 8133 (1990).

\bibitem{BW2} B. Blok and X. -G. Wen, Effective theories of the fractional quantum Hall effect: Hierarchy construction, {\it Phys. Rev. B }{\bf 42}, 8145 (1990).

\bibitem{XA} X. Artru, String model with baryons: Topology; classical motion, {\it Nucl. Phys. B }{\bf 85}, 442 (1975).

\bibitem{IP} N. Isgur and J. Paton, Flux-tube model for hadrons in QCD, {\it Phys. Rev. D }{\bf 31}, 2910 (1985).

\bibitem{AN} O. Andreev, Some multiquark potentials, pseudopotentials, and AdS/QCD, {\it Phys. Rev. D }{\bf 78}, 065007 (2008).

\bibitem{NJL} Y. Nambu and G. Jona-Lasinio, Dynamical model of elementary particles based on an analogy with superconductivity. I, {\it Phys. Rev. }{\bf 122}, 345 (1961).

\bibitem{SA} J. J. Sakurai, Vector-meson dominance and high-energy electron-proton inelastic scattering, {\it Phys. Rev. Lett. }{\bf 22}, 981 (1969).

\bibitem{THO} G. ’t Hooft, Magnetic monopoles in unified theories, {\it Nucl. Phys. B }{\bf 79}, 276 (1974).

\bibitem{POL} A. M. Polyakov, Particle spectrum in quantum field theory, {\it JETP Lett }{\bf 20}, 194 (1974).

\bibitem{SKY} T. H. R. Skyrme, A unified field theory of mesons and baryons, {\it Nucl. Phys. }{\bf 31}, 556 (1962).

\bibitem{WIT} E. Witten, Current algebra, baryons, and quark confinement, {\it Nucl. Phys. B }{\bf 79}, 223 (1983).

\bibitem{HEST1} D. Hestenes,  {\it  Space-Time Algebra}, (Gordon and Breach, New York, 1966).

\bibitem{HEST2} D. Hestenes and G. Sobczyk,  {\it  Clifford algebra to geometric calculus: a unified language for mathematics and physics}, (Kluwer Academic Publishers, Dordrecht, 1984).

\bibitem{PAV} M. Pavsic,  {\it The Landscape of Theoretical Physics: A Global View. From Point Particles to the Brane World and Beyond, in Search of a Unifying Principle}, (Kluwer Academic Publishers, Dordrecht, 2001).

\bibitem{Loun} P. Lounesto,  {\it Clifford algebras and spinors}, (Cambridge University Press, Cambridge, 2001).

\bibitem{DORA} C. Doran and A. Lasenby,  {\it  Geometric Algebra for Physicists}, (Cambridge University Press, Cambridge, 2003). 

\bibitem{WB} W. E. Baylis, {\it Clifford (Geometric) Algebras: with applications to physics, mathematics, and engineering}, (Springer Science \& Business Media, Boston, 2012).

\bibitem{WL1} W. Lu, Yang-Mills interactions and gravity in terms of Clifford algebra, {\it Adv. Appl. Clifford Algebras } {\bf 21}, 145 (2011).

\bibitem{WL2} W. Lu,  Electroweak and Majorana Sector Higgs Bosons and Pseudo-Nambu-Goldstone Bosons, arXiv:physics.gen-ph/1603.0469.

\bibitem{WL3} W. Lu, A Clifford algebra approach to chiral symmetry breaking and fermion mass hierarchies, {\it Int. J. Mod. Phys. A }{\bf 32}, 1750159 (2017).

\bibitem{WL4} W. Lu,  Modified Einstein-Cartan gravity and its implications for cosmology, arXiv:gr-qc/1406.7555.

\bibitem{WL5} W. Lu, Dynamical symmetry breaking and negative cosmological constant, {\it Int. J. Mod. Phys. A }{\bf 34}, 1950136 (2019).

\bibitem{HEST3} D. Hestenes,  Gauge gravity and electroweak theory, {\it Found. Phys. } {\bf 12}, 153 (1982).

\bibitem{PAV2} M. Pavsic, Spin gauge theory of gravity in Clifford space: A realization of Kaluza-Klein theory in 4-dimensional spacetime, {\it Int. J. Mod. Phys. A }{\bf 21}, 5905 (2006).

\bibitem{TRAY} G. Trayling and W. E. Baylis, A geometric basis for the standard-model gauge group, {\it J. Phys. A }{\bf 34}, 3309 (2001).

\bibitem{HA} B. I. Halperin, Statistics of quasiparticles and the hierarchy of fractional quantized Hall states, {\it Phys. Rev. Lett. }{\bf 52}, 1583 (1984).

\bibitem{ASW} D. Arovas, J. R. Schrieffer and F. Wilczek, Fractional statistics and the quantum Hall effect, {\it Phys. Rev. Lett. }{\bf 53}, 722 (1984).

\bibitem{LV} LIGO Scientific and Virgo Collaboration, (B. P. Abbott {\it et al.}), GW170817: observation of gravitational waves from a binary neutron star inspiral, {\it Phys. Rev. Lett. }{\bf 119}, 161101 (2017).

\end{thebibliography}
\end{document}